\documentclass[a4paper,fleqn,usenatbib]{mnras}

\usepackage[T1]{fontenc}
\usepackage{ae,aecompl}

\usepackage{graphicx}	
\usepackage{amsmath}	
\usepackage{amssymb}	

\title[Hot-Jupiters formation location]{Disentangling Hot Jupiters formation location from their chemical composition}

\author[M. Ali-Dib]{
Mohamad Ali-Dib$^{1}$$^{,}$$^{2}$\thanks{E-mail: m.alidib@utoronto.ca}
\\
$^{1}$Centre for Planetary Sciences, Department of Physical \& Environmental Sciences, University of Toronto at Scarborough,\\
Toronto, ON M1C 1A4, Canada\\
$^{2}$Canadian Institute for Theoretical Astrophysics, 60 St. George St, University of Toronto, Toronto, ON M5S 3H8, Canada\\
}

\date{Accepted XXX. Received YYY; in original form ZZZ}

\pubyear{2016}

\begin{document}
\label{firstpage}
\pagerange{\pageref{firstpage}--\pageref{lastpage}}
\maketitle

\begin{abstract}
We use a population synthesis model that includes pebbles and gas accretion, planetary migration, and a simplified chemistry scheme to study the formation of hot-Jupiters. Models have been proposed that these planets can either originate beyond the snowline and then move inward via disk migration, or form ``in-situ'' inside the snowline. The goal of this work is to verify which of these two scenarios is more compatible with pebble accretion, and whether we can distinguish observationally between them via the resulting planetary C/O ratios and core masses. Our results show that, for solar system composition, the C/O ratios will vary but {moderately} between the two populations, since a significant amount of carbon and oxygen are locked up in refractories. In this case, we find a strong correlation between the carbon {and oxygen} abundances and core mass. The C/O ratio variations are more pronounced in the case where we assume that all carbon and oxygen are in volatiles. On average, Hot-Jupiters forming ``in-situ'' inside the snowline will have higher C/O ratios because they accrete less water ice. However, only Hot-Jupiters forming in-situ around stars with C/O=0.8 can have a C/O ratio higher than unity. We finally find that, even with fast pebble accretion, it is significantly easier to form Hot-Jupiters outside of the snowline, even if forming these ``in-situ'' is not impossible in the limit of the simplifying assumptions made.

\end{abstract}



\section{Introduction}
Since their first detection in 1995 \citep{mayor}, the origin of Hot Jupiters has been covered with mysteries. These were first thought to originate beyond the water iceline were solid material is abundant enough for their massive cores to form via classical planetesimals accretion \citep{pollack1996}, then migrate inward via either disk migration \citep{goldreich,ward,kley2012} or high eccentricity migration \citep{rasio,wu}. This paradigm got challenged recently with many new theoretical and observational advancements. First, pebble accretion was introduced as a very efficient mechanism for cores formation \citep{lamb2}, opening a new possibility to form these even inside the water snowline. Second, newly found extreme planetary systems similar to WASP 47 \citep{wasp471,wasp472} with a very fragile architecture (non hierarchical tightly packed multiple massive planets including a Hot Jupiter) puts into question any migration scenario for these systems, since this would probably destabilize them. For these reasons, the possibility of in-situ formation some of these planets has been recently invoked \citep{hjin1,hjin2}. This hypothesis however suffers from its own problems. First and for all it is still unclear what would stop the planets forming in-situ from spiraling into the central star, but this is a problem faced also by planets forming just beyond the snowline. The other problem for in-situ Hot Jupiters formation is their still relatively long formation timescale in this region where pebbles are smaller and the very high atmospheric temperature along with the low {pebble isolation mass}, severely increasing the contraction timescale for their envelopes \citep{bitsch1}. For all these reasons, an observational method or criteria should be developed and used to disentangle between these very different formation scenarios. The carbon and oxygen chemical abundances, and their C/O ratio, have long been proposed as tracers for planets formation processes from both theoretical (\cite{oberg,madhu2,ali-dibb} \cite{alibert1,morda2016}) and observational \citep{madhu1,kreid1,kreid2,steve,line} point of views (we refer the readers to \citep{madhu3} for a recent review on both aspects). These species are present throughout the disk in partly volatile phases that condense when temperatures are low enough, allowing cores forming beyond a certain element's snowline to be rich in the said element. Linking the chemical composition of a giant planet to its formation location and mechanism is tricky though since a large number of processes are at play. A global model incorporating simplified versions of these processes is hence needed to quantify the chemical composition of a giant planet as a function of its initial formation conditions, which is the subject of this work. The questions we will try to answer are: 
\begin{itemize}
\item Can planetary seeds inside the water snowline grow into Jupiter mass planets before the dissipation of the disk ?
\item Can Hot Jupiters forming in situ be distinguished from those who migrated from outside the snowline via their chemical composition ?
\item {What is the impact of the disk's chemistry on the composition of a giant planet, and what is the best chemical tracer for formation processes ?}
\item What is the effect of the formation location on the total mass of heavy elements in the planet ? 
\end{itemize}

In section 2 we introduce the model we use to connect the planet's chemistry to its formation. In section 3 we present the results and we conclude in section 4.

\section{Model}
The model we use in this work is similar to the one used in \cite{ali-dibc}, based on \cite{lamb1,lamb3,bitsch1,bitsch2}. This model includes the following:
\begin{itemize}
\item Fits to a 2D disk model {(on a discrete grid)} with accurate opacities and radiative treatment, but not viscous evolution. {The opacity model is important for its effects on the thermal structure of the disk (where the dust opacity regulate the temperature profile, and by consequence pressure and density) and thus migration due to its dependence on the density and temperature gradients.  }
\item Parametric pebbles \& gas accretion. pebble accretion is much more efficient in forming planetary cores than the classical planetesimals accretion \citep{ormel}, with the pebbles size and surface density being the fundamental parameters. {Pebbles are assumed to form \& replenish continuously from coagulation and collisions beyond the ``pebbles production line'' in the outer disk.} The gas accretion depend dominantly on the envelope's opacity, density and temperature, in addition to the core's mass.   
\item Type I and II migration through torques evaluation. Type I migration will affect low mass planets through the Lindblad and corotation torques, while type II migration will affect planets massive enough to open a gap in disk and follow its viscous evolution. We assume that the planets inward migration will stop at the inner cavity, and hence won't be lost to the star. Disk migration is implemented in all simulations, so when discussing Hot-Jupiters forming ``in-situ'', we mean inside of the water snowline. 
\item The model tracks the carbon and oxygen abundances of the forming planet. These elements are each divided into refractory and volatiles phases. The refractory phase is solid at all temperatures, and the volatile part is in ice phase only outside of the snowline. We assume that all carbon and oxygen are in water and CO (or CO$_2$). The refractory to volatile ratio is a free adjustable parameter. We study a solar system refractory/volatile ratio case in addition to a purely volatile case. {For simplicity, we assume that the refractories mass is dominated by C and O, and we neglect contributions from other elements, like Fe and Si.}
\item A planet's core can get eroded {if the temperature of the core-envelope boundary is higher than its thermodynamical stability limit. Convection can then pollute the envelope with heavy elements {from this core \citep{stevenson1982,wm1,wm2,guillot2004}.}} We treat the core's erosion through a free parameter ranging from 0\% (no core erosion) to 100\% (full core erosion). {We refer the reader to \cite{ali-dibc} for a discussion on {this approach}.}  
\end{itemize}

\begin{figure*}
	\includegraphics[scale=0.1]{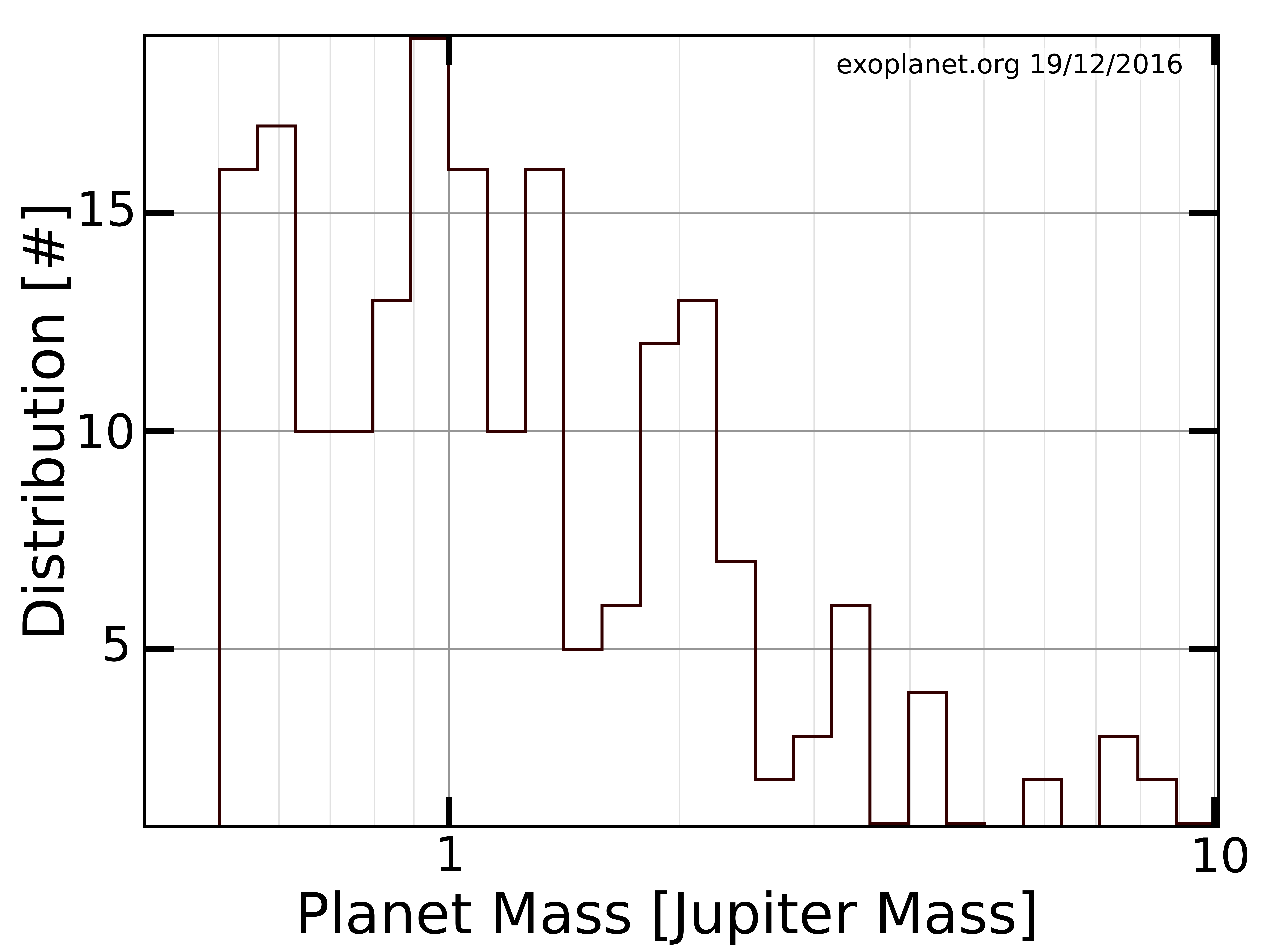}
   \caption{{The mass distribution of all observed {Hot-Jupiters}, from \texttt{Exoplanets.org}.}}
    \label{fig:referee}
\end{figure*}

{Simulations were run on a 0.025 AU spaced grid. Therefore the first disk grid point is at 0.025 AU from the star, which is the reason why we set this point as our inner disk limit, and also the reason why the ``Final positions'' in all plots are discrete (this is more visible in the inner disk since a log scale was used in these plots). We stop the simulations when a planet reaches this point because it is uncertain if its migration will stop or if it will get accreted to the star. {The goal of this work is to understand the chemistry of giant planets (in term of C/O ratio), not their physical characteristics. For this reason, and to make the simulations numerically faster, we artificially stopped the simulations when a planet reached 1 Jupiter mass. This is Justified by Fig. \ref{fig:referee} from \texttt{exoplanets.org} showing that 1 Jupiter mass planets are the most common among {Hot-Jupiters, with this distribution decreasing steadily for heavier planets. Moreover, as discussed in section 3.1, the measured chemical composition of the envelope scale very slowly with its mass, assuming significant core erosion.}} We additionally excluded any planet below 100 M$_\oplus$ from the analysis so no Super-Earths are present. Therefore, even if this approximation is not exact, it is justified and reasonable for the purpose of the paper. Since we are studying the C/O ratio of planets, the effect of higher amounts of gas in the envelope is very limited, and will almost not affect the C/O of a giant planet (as discussed below). Throughout this paper, we define a Hot-Jupiter as planets heavier than 100 M$_\oplus$, inside 0.1 AU.}

For a more detailed description of the model we refer the reader to \cite{ali-dibc} and the references therein.
However, the model in \cite{ali-dibc} was developed primarily with the solar system's giant planets in mind. Modifications should be made to account for the differences related to hot-Jupiters. If hot-Jupiters formed in-situ (inside the water snowline), their initial planetary seeds would have encountered a pebbles disk structure significantly different than beyond the snowline. As \cite{morby1} have noted, the pebbles inward flux {(equation 4 in \cite{ali-dibc})} might be 2 times lower inside the snowline, with the pebbles characteristic size an order of magnitude smaller. We take this into account by artificially lowering the mass flux and pebbles characteristic size inside the water snowline by these factors {of respectively 2 and 10. The factor 2 reduction in the pebbles mass flux is consistent with our nominal case solar system chemistry where 50\% of water is in refractories}.

Such a global model usually include a large number of free parameters. To keep the problem tractable, we only vary the parameters that are assumed to affect directly the planet's chemical compositions. The free parameters space is explored through population synthesis, {as shown in tables \ref{t1} and \ref{t2}}. {These parameters are: T$_{ini}$ and R$_0$ (the seed's injection time and location), E$_f$ (the core's erosion factor), M$_0$ (the seed's initial mass), ``metal'' (the disk's metallicity in small coupled dust grains) and Z$_0$ (the disk's metallicity in large decoupled pebbles). We fix the value of the following parameters in all simulations: $f$ (a fudge factor that reconciles our simplified slow phase gas accretion rate parametric fit with more detailed hydrodynamic simulations \citep{piso}), $\kappa_{env}$ (the envelope opacity), and $\rho_c$ (the core's density).}\\

\section{Results \& Discussions}
\subsection{Analytical discussion \& global properties}
\label{analytic}
Even though numerical simulations are needed to extract precise quantitative predictions from the model, its basic aspects can be understood from simple arguments. The model takes into account two possible sources for carbon and oxygen: the eroded core and the envelope. During the envelope's hydrodynamical collapse, the disk's gas along with {all of the solids (grains and pebbles) it contains,} will be accreted, leading to a necessarily stellar C/O ratio (assuming that any dust accreted with the envelope will sublimate and mix with the hot gaseous envelope.) even if this quantity has evolved in the separate gas and solid phases \citep{oberg}. The source for any deviation in  the planet's C/O ratio from the stellar value should hence originate from the partially or completely eroded core. Therefore if the core has a stellar C/O ratio, the bulk observable planet's value will remain intact. If on the other hand the core had a super or substellar C/O ratio, then (assuming significant core erosion) the envelope's C/O will follow in the same direction. \\
{Before moving forward with the discussion, let's check the magnitude of the core erosion's effect on a giant planet's envelope using some simple back of the envelope calculations:\\
Assume a simple planet {forming entirely inside the water iceline} with 10 M$_\oplus$ {completely refractory} core, {that is 40\% in mass Carbon and 60\% in mass Oxygen (so molar C/O of 0.88, cf. table \ref{t2})}, {with a 300 M$_\oplus$ envelope of solar composition (so C/O=0.55)}, and let's assume 100\% core erosion.
The measured final total planet C/O ratio will be:\\
\begin{equation}
C/O_{tot}=\frac{(C_{erosion}+ C_{envelope})}{(O_{erosion} + O_{envelope})}
\end{equation}
where C$_{erosion}$ and C$_{envelope}$ are respectively the core and envelope contribution to the final planet C/O ($C/O_{tot}$).
We can write 
\begin{equation}
C_{erosion}=0.4\times M_{core}/(Mol_C)
\end{equation}
where $M_{core}$ is the core total mass (before erosion), $(Mol_C)$ is the Carbon's molar mass, and factor {0.4 is introduced because the core is 40\% in mass carbon.}
and :
\begin{equation}
C_{envelope}=0.55\times(10^{-3})*(M_{atm})/{Mol_{H_2}}
\end{equation}
where $M_{atm}$ is the core atmosphere mass (before erosion), $(Mol_{H_2})$ is the Hydrogen's molar mass, {10$^{-3}$ is the Oxygen's relative molar abundance (in solar proportions) \citep{asplund}} and 0.55 the C/O ratio.
Putting all these numbers together we get finally C/O$_{tot}$ of {around 0.8}. So a fully eroded 10 M$_\oplus$ core can shift the {planet's total average C/O ratio toward the core's value}. Moreover, in the limit of very large Envelope mass (order of magnitude more massive than Jupiter), this formula will give a C/O of 0.55, the solar value. 
}

\begin{figure*}
	\includegraphics[scale=0.35]{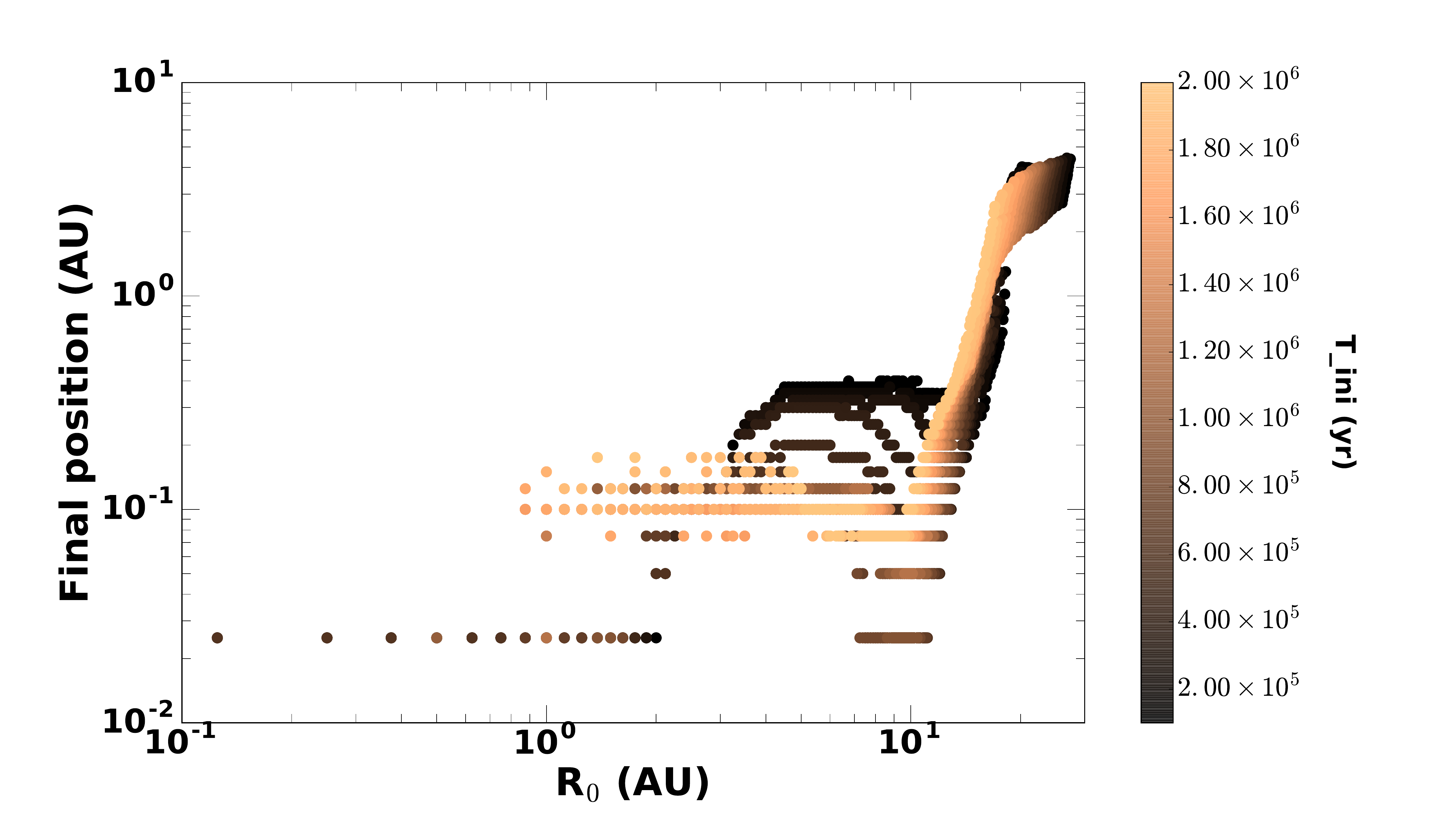}
   \caption{The final position (Rf) of all Jupiter mass planets as a function of the seed's initial position (R0) and injection time (T$_{ini}$) with no photoevaporation included.}
    \label{fig:alljups}
\end{figure*}

The core C/O depend on two quantities: the amount of accreted volatile ices and the amount of accreted refractory species. A superstellar C/O ratio cannot be achieved from purely volatile ices, simply because the CO and CO$_2$ icelines are both further out in the disk than the water's snowline. There is hence no way for a core to accrete more C-bearing ices than water ices (in the framework of the classical core accretion scenario, but cf. \cite{ali-dibb}). A planet accreting volatiles ices as the unique source of C and O will hence have strictly stellar or substellar C/O ratios. We can conclude therefore that an important refractory carbonaceous component is necessary for a hot Jupiter to have a superstellar C/O ratio. This is however not the only constraint. {For this to happen, the planet also necessarily need to accrete more carbon than oxygen, and since all refractories are solid at all temperatures, the only way for a planet to obtain this superstellar C/O ratio is for the refractory dust to be more rich in carbon than oxygen. This seems to be the case in our own solar system from Halley's in situ measurements \citep{jessberger1988}, even though this is not settled.} \\
Figure \ref{fig:alljups} shows the population of all ``Jupiters'' (defined as 100 M$_\oplus$ $\leq$ M $\leq$ 320 M$_\oplus$) found in the simulations. It shows the final positions of the planets as a function of the seeds initial placement and injection time into the disk. We can find in this population cold, warm and hot Jupiters. More interestingly, we find a population of Hot Jupiters who form in the inner most parts of the disk (``in situ''). We notice that all of the Hot Jupiters parked at the inner most edge of the disk and who started forming inside the snowline have started forming very early in the disk (2.5$\times$ 10$^5$ yr). This is expected since only that early in the disk's lifetime the seed will have time to grow large via the slow Bondi regime to start accreting pebbles in the much faster Hill regime. Additionally, only in young disks the pebbles isolation mass is high enough to allow the formation of a core massive enough to contract an atmosphere before the dissipation of the disk. This shows that the formation of Hot Jupiters ``in situ'' under the multiple assumptions present in the model is possible, but it should take place very early in the disk. However, again by looking at Figure \ref{fig:alljups}, we notice that most of the Hot-Jupiters found by this model started forming in the outer disk. So while it is possible to form Hot-Jupiters in situ, it is much easier and more efficient to form them starting in the outer disk.

\subsection{Case: solar chemistry}
\begin{figure*}
	\includegraphics[scale=0.35]{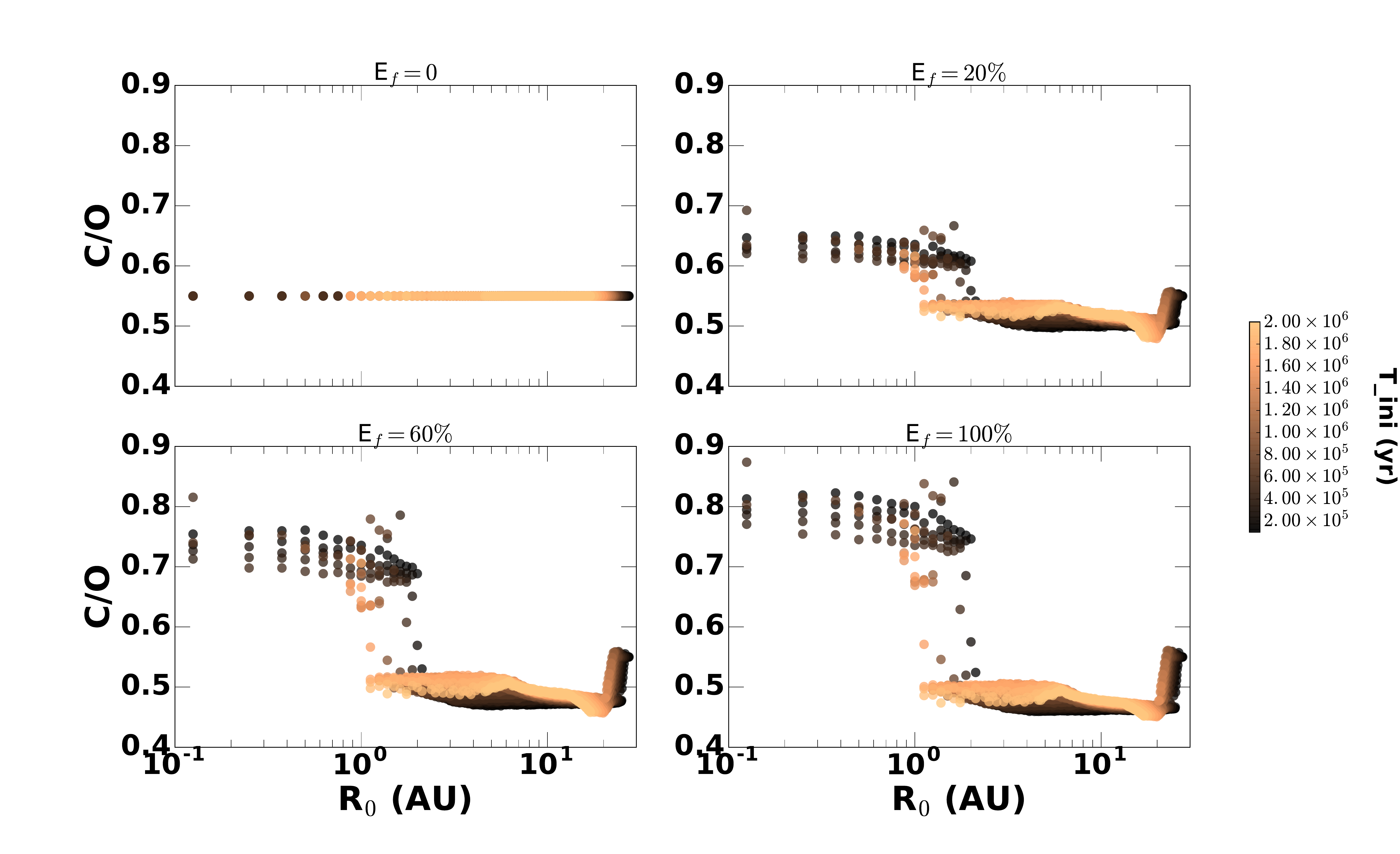}
   \caption{The C/O of Hot Jupiters as a function of the seed's initial position (R0) and injection time (T$_{ini}$) in the case with no disk photoevaporation and stellar C/O = 0.55. The four cases are for different core erosion factors. The C/O difference between planets forming inside and outside of the snowline grow wider with the core erosion factor since more accreted ices get dissolved into the envelope.}
    \label{fig:jup1}
\end{figure*}

\begin{figure*}
	\includegraphics[scale=0.30]{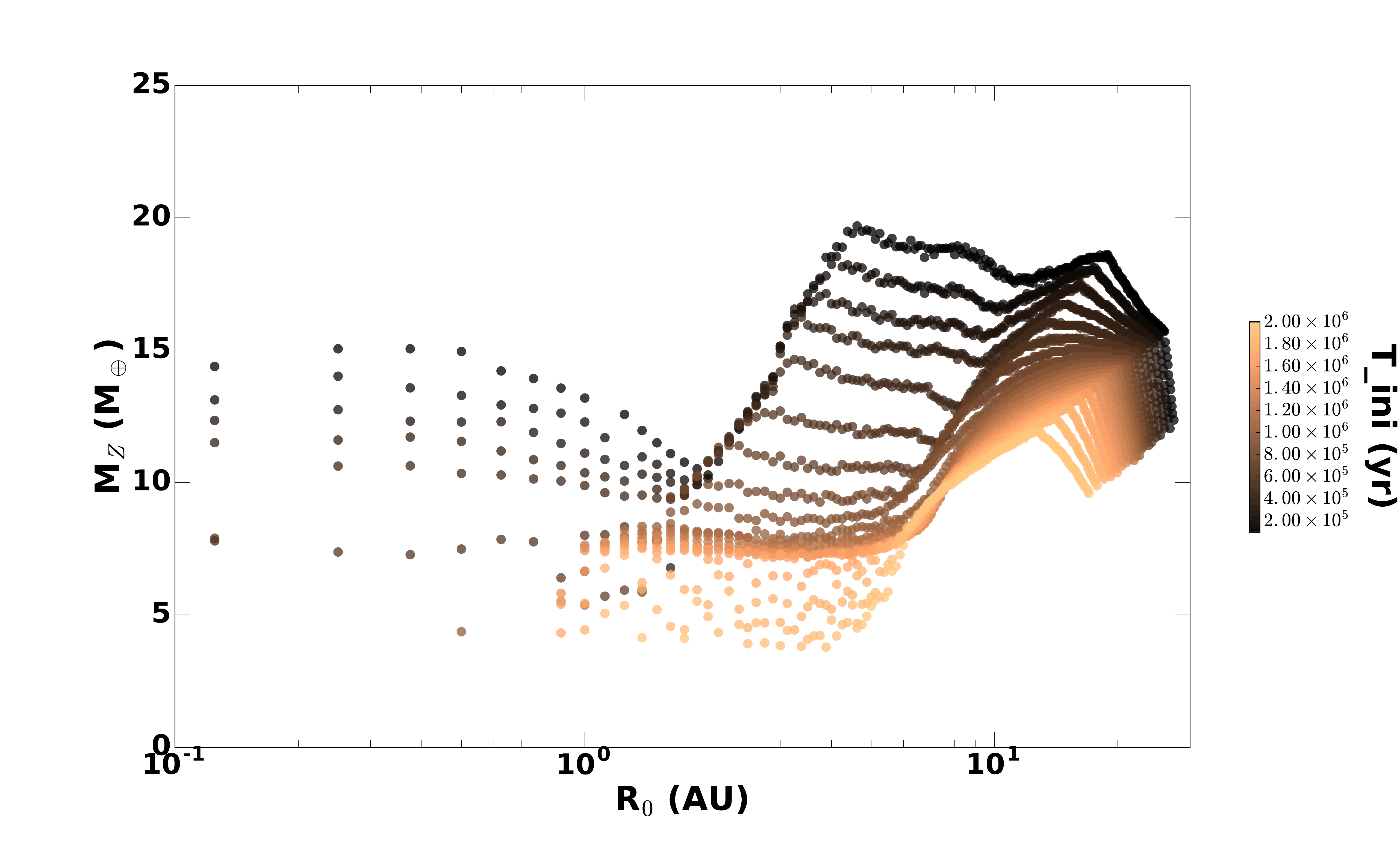}
   \caption{The total mass of heavy elements in the core and envelope ($M_Z$) of Hot Jupiters as a function of their starting formation location and time, assuming completely eroded cores. Planets forming early and far out in the disk will end up with higher $M_Z$ due to their higher pebbles isolation mass.}
    \label{fig:mznope}
\end{figure*}

We first set a solar C/O ratio with solar refractory/volatile ratios. Figure \ref{fig:jup1} show the C/O ratios of Hot Jupiters resulting from the population synthesis as a function of their seed's initial injection place and time, and for different core erosion factors. For a Hot Jupiter forming between the water and CO icelines, the core will be made from both refractory and icy oxygen phases, but only refractory carbon. In this case, with core erosion, more oxygen will get ceded into the envelope than carbon, and thus the erosion will slightly decrease the planet's C/O ratio (because both carbon and oxygen in this case have very abundant refractory phases). If however the planet's core formed mainly beyond the CO iceline, it will be accreting all of the chemical species in solid phase, and hence will maintain its solar C/O ratio. The envelope will hence maintain its C/O ratio no matter the amount of core erosion. In the third case where the planet forms entirely inside the water's snowline, its C/O ratio will depend on the amount of these species present in refractory dust. Since in our case there is more carbon refractories than oxygen refractories, the planet's C/O will increase. All these different cases can be seen in Figure \ref{fig:jup1}, where we find that planets forming inside the water snowline have C/O ratio up to 0.8, while those that formed between the two snowlines have a slightly subsolar C/O of around 0.45, and finally the planets who formed all the way far out beyond the CO iceline have solar C/O value.\\
Figure \ref{fig:mznope} shows the total mass of heavy elements in the hot Jupiter ($M_Z$) as a function of their seed's initial injection place and time {in the case of complete core erosion}. The values we find are in line with the indirect retrieval method of \cite{thorn}. Planets forming earlier and further out in the disk have higher isolation masses and thus higher $M_Z$\footnote{Since the core's isolation mass scales with the aspect ratio (H/r) of the disk.}. We note however the degeneracy between injection place and time, making this criteria on its own unconstraining. It needs to be coupled to a chemical abundance measurement for useful informations on a planet's history. \\

\begin{figure*}
	\includegraphics[scale=0.30]{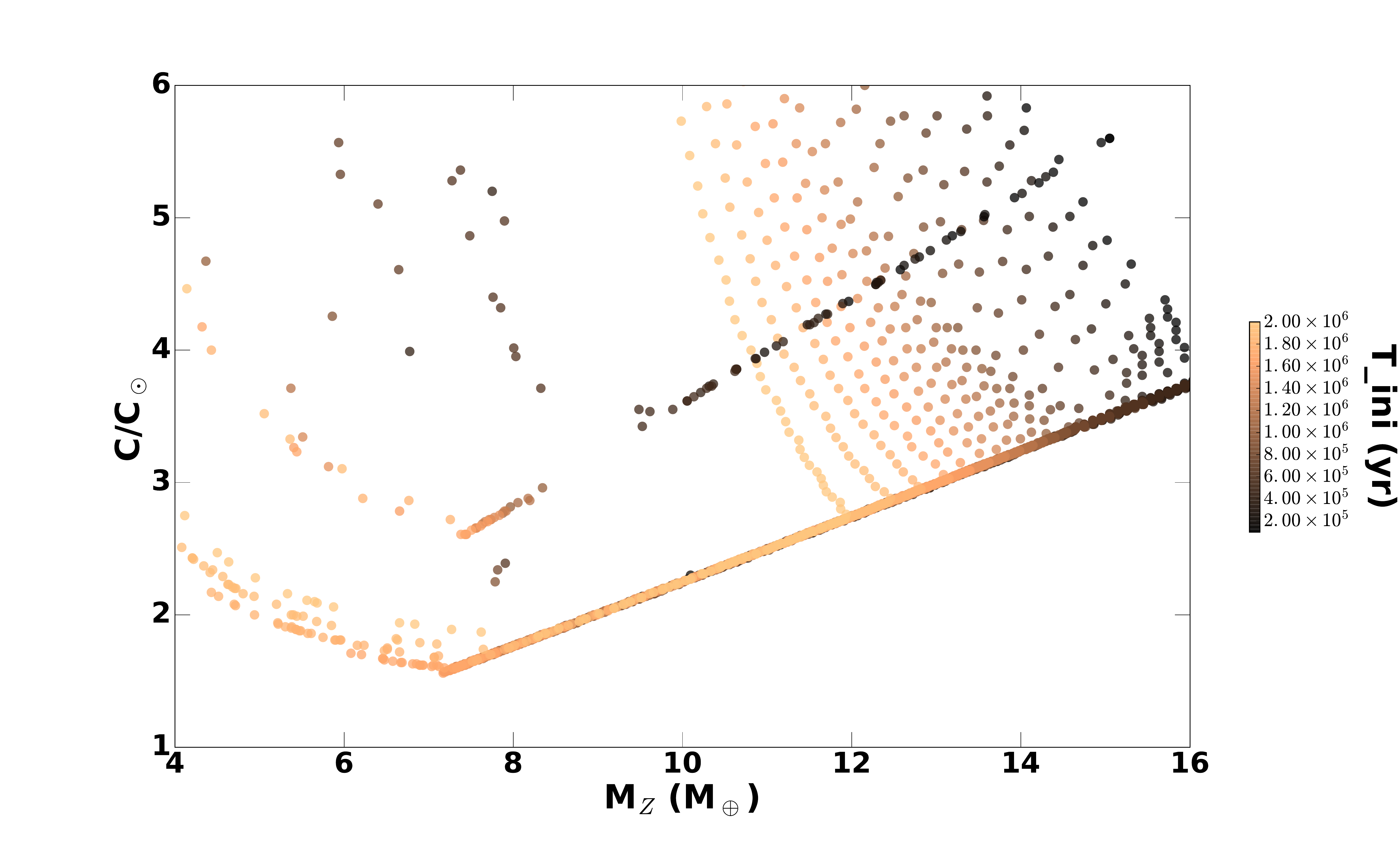}
   \caption{The carbon abundance enrichment (normalized with respect to the stellar value) as a function of the total mass of heavy elements in the core and envelope ($M_Z$) of Hot Jupiters, assuming completely eroded cores.}
    \label{fig:cmz}
\end{figure*}

\begin{figure*}
	\includegraphics[scale=0.30]{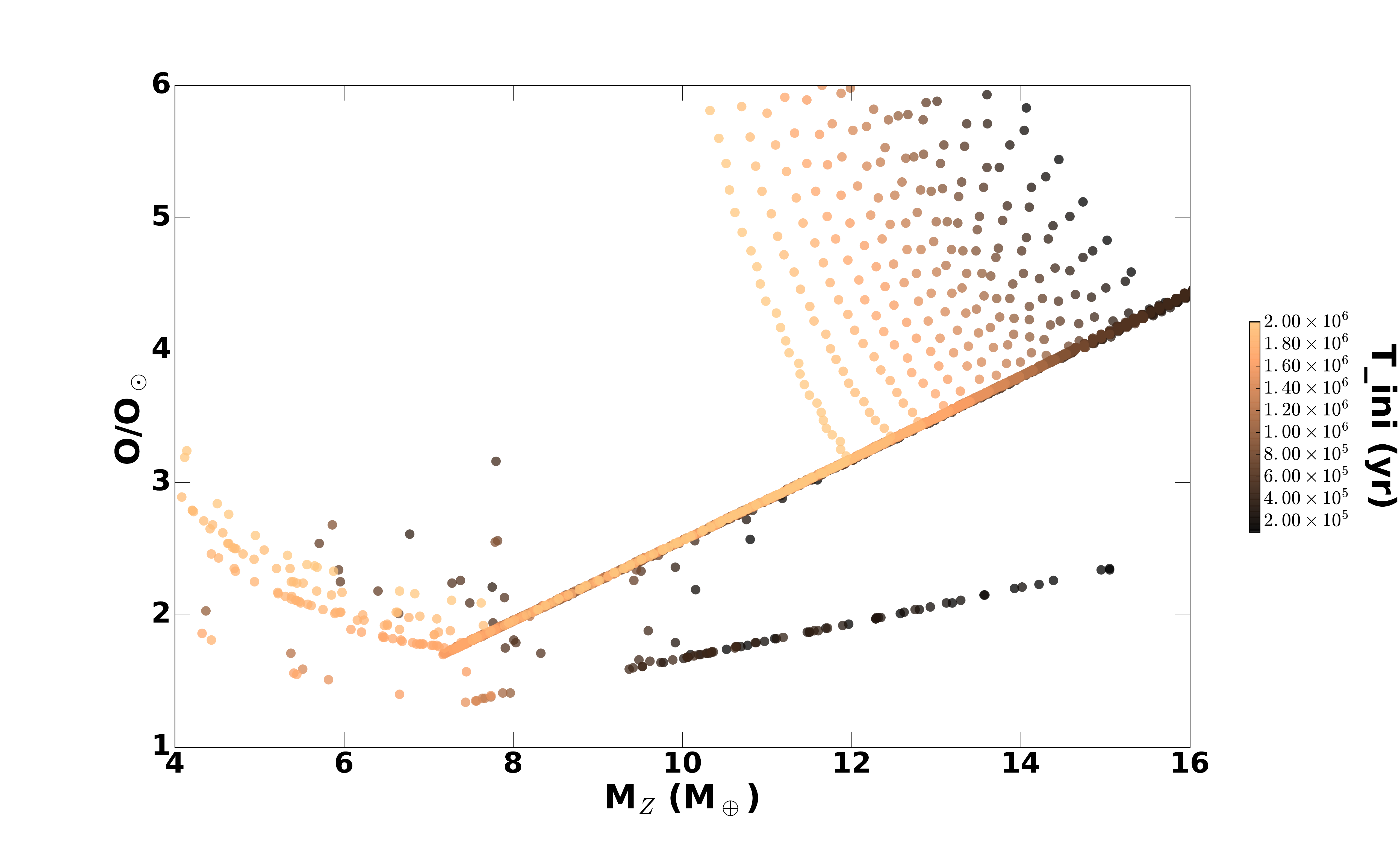}
   \caption{The oxygen abundance enrichment (normalized with respect to the stellar value) as a function of the total mass of heavy elements in the core and envelope ($M_Z$) of Hot Jupiters, assuming completely eroded cores.}
    \label{fig:omz}
\end{figure*}

{Figures \ref{fig:cmz} and \ref{fig:omz} show respectively the planet's bulk carbon and oxygen enrichments as a function of $M_Z$, where {linear correlation trends are seen in each case between the two observables.} {We notice that both C and O abundances are always above solar value simply because we are only taking the fully eroded core case, and thus an enrichment above solar abundance is always present in the envelope.} Oxygen and/or carbon abundances measurements can hence be used to constrain $M_Z$, or the other way around. This method is in analogy with \cite{ali-dibc} method of constraining $M_Z$ and the core masses of Jupiter and Saturn from sulfur and phosphorus observations.\\ The values retrieved using this method should be considered lower limits however due to the uncertainty on the core erosion factors in Hot Jupiters and the amount of carbon and oxygen in refractories. {In other words, for any planet, we do not know if we are measuring the entire carbon/oxygen budget of the planet (fully eroded/well mixed) or just a small fraction that was eroded into the envelope. For this reason, any total mass of heavy elements of the planet inferred from this correlation should be treated as a lower limit since we do not know for sure if we are observing the entire carbon inventory or just a small fraction of it. \\
Two separate trends can be seen for both elements, the upper oxygen and lower carbon branches corresponding the Hot Jupiters who formed outside of the water snowline through migration, and the lower oxygen and upper carbon branches corresponding to the ``in-situ'' Hot Jupiters not accreting water ice, but only oxygen and carbon in refractories.
In the 2 main branches (forming outside the water iceline) we distinguish 2 different populations: The straight line represents the planets that formed between the water and CO icelines, while {the swarm on the upper right beyond 11 AU are the planets whose cores were accreted partially beyond the CO iceline, and accreted more carbon.}}
The two separate trends for the two possible formation location for Hot-Jupiters lead to a degeneracy in the values of $M_Z$ for the same carbon and oxygen abundances between {planets forming early and entirely inside the iceline, and those forming late and far outside of the iceline.} Using both chemical indicators simultaneously can possibly remove this degeneracy.}

\subsection{Case: pure volatiles}
\begin{figure*}
	\includegraphics[scale=0.30]{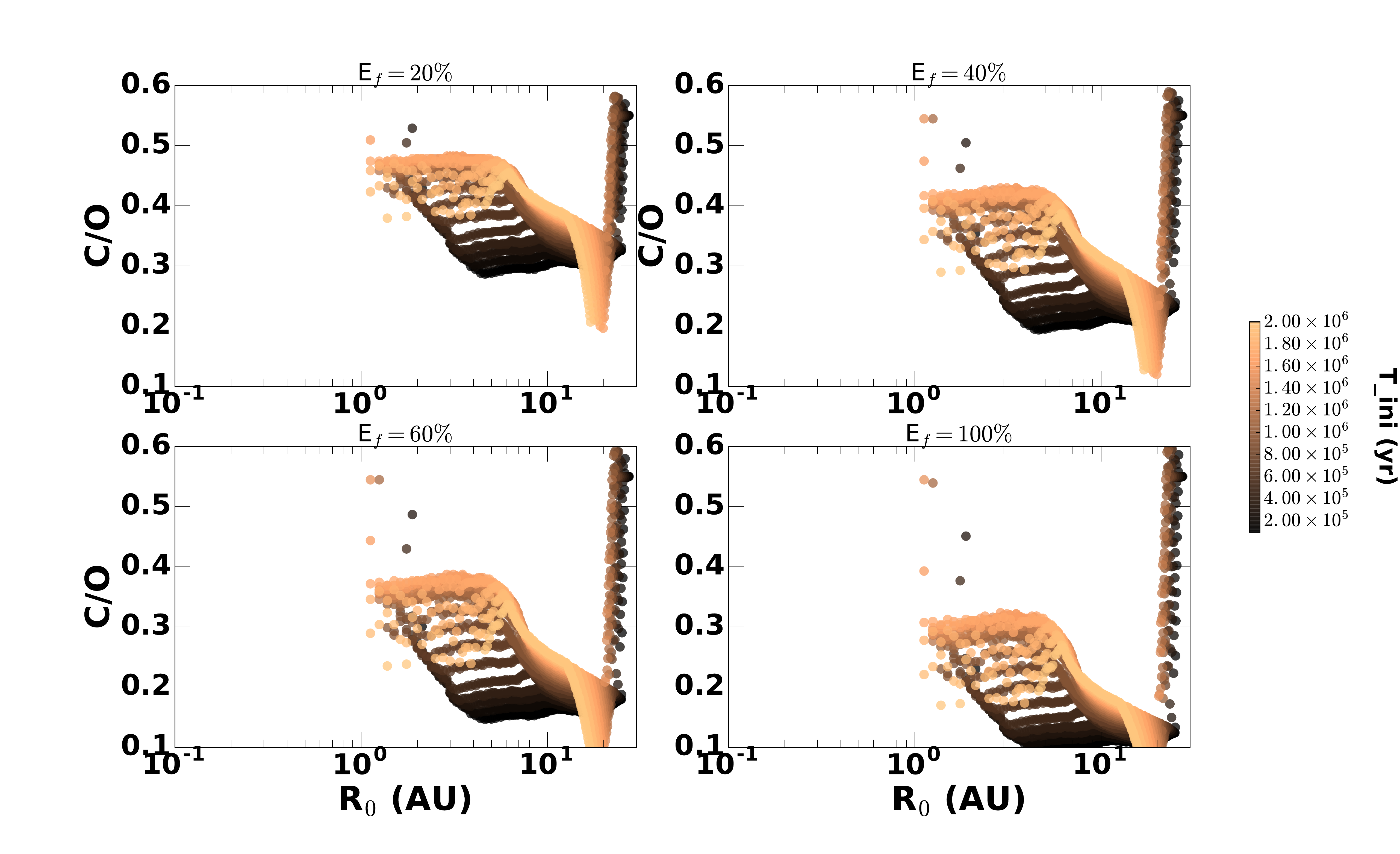}
   \caption{Same as Fig. \ref{fig:jup1} but with all elements in purely volatile phases. {Here we show the 20\% and 40\% core erosion cases instead of 0\% that should be identical to the solar composition case (C/O = 0.55 for all planets). In these simulations, no planet can form inside of the water iceline. } }
    \label{fig:noref}
\end{figure*}

\begin{figure*}
	\includegraphics[scale=0.30]{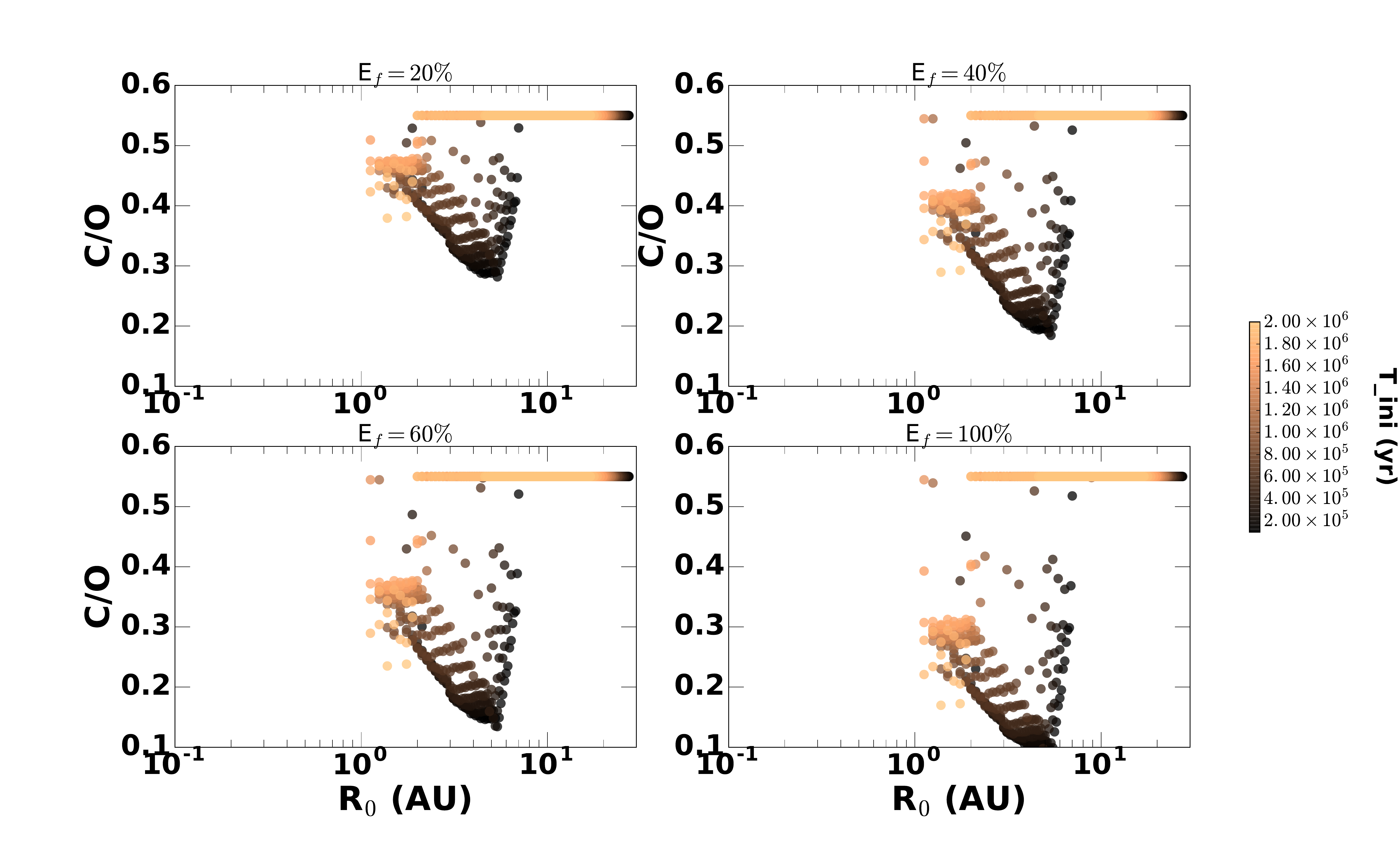}
   \caption{Same as Fig. \ref{fig:noref} but with all carbon in CO$_2$ instead of CO. }
    \label{fig:co2}
\end{figure*}

We now run simulations {in an extreme scenario where all the elements outside of the water snowline are }in the purely volatile phase (no carbon or oxygen in refractories). {The goal of this case is to demonstrate the effect of possible extreme volatile/refractory ratios.} The final C/O distribution should hence be even more dependent on the initial position of the planet since the total amount of accreted carbon and oxygen will depend entirely on the planet's position with respect to the snowlines. {In this extreme case, in the absence of refractories, no solids at all are present inside the water iceline. We hence only run simulations with cores forming outside of this location.}\\
 The results are shown in Figure \ref{fig:noref}. We notice that the obtained planets follow a broad distribution where planets forming inside the water snowline and outside of the CO's iceline have solar C/O while those forming in between show much lower values depending on the amount of ices accreted. This is understandable since only between the two icelines oxygen fractionates (in different physical phase) from carbon.  In this case however The C/O can be as low as 0.1 for planets that did not accrete any CO ices while accreting important amount of water ices, leading to a slightly steeper distribution trend. This case however is for illustration only. It is unlikely that a disk can exist with purely volatile phases. We know from small bodies observations that this was far from the reality in our own protosolar nebula, and observations in TW Hya and HL Tau show a strong depletion in CO with respect to the canonical value, hinting toward the CO transforming into other elements, either CO$_2$ or refractories \citep{co1,co2,co4}.\\
We explore this in Fig. \ref{fig:co2} showing the results from simulations where all of the carbon is in CO$_2$ ({condensing at 80 K}). In this case, similar to the CO case, all planets forming inside the water iceline or outside of the CO$_2$ iceline will maintain solar C/O ratio, while only planets forming between these two will deviate. Interestingly, since CO$_2$ condense at significantly higher temperatures than CO, the CO$_2$ iceline can move significantly inward to the inner disk during the disk's lifetime. Hence planets who form outside of this iceline (hence maintain the solar C/O ratio) late enough will be located in the inner most parts of the disk.

\subsection{Case: High stellar C/O ratio}
\begin{figure*}
	\includegraphics[scale=0.35]{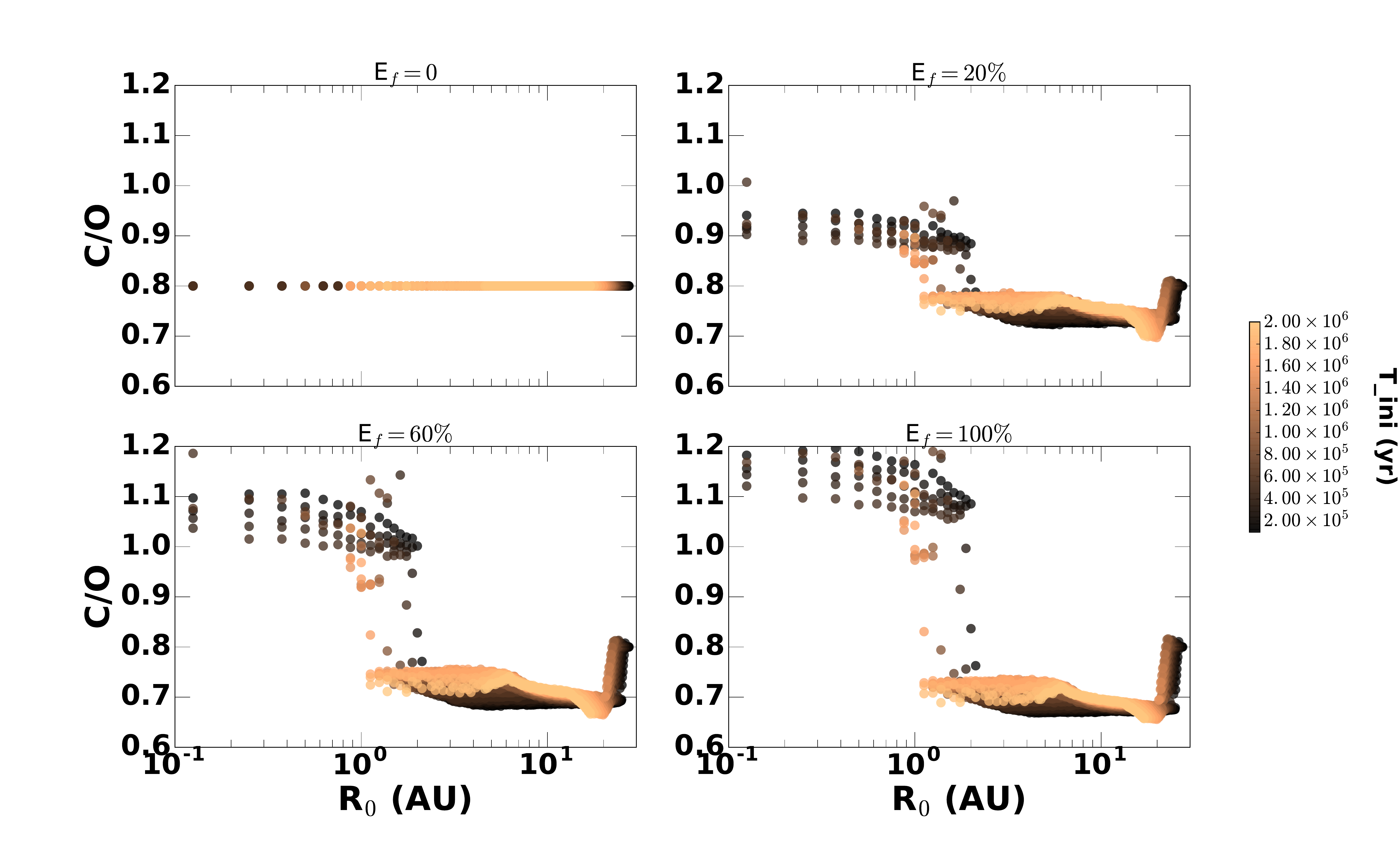}
   \caption{Same as Fig. \ref{fig:jup1} but for stellar C/O = 0.8}
    \label{fig:jup2}
\end{figure*}

Ultimately, the fundamental parameter affecting the planets chemistry is the host star's C/O ratio (imprinted onto the disk). Multiple surveys studied the distribution of the stellar population C/O ratios \citep{stellar1,stellar2,stellar3,brewer}. These works concluded that planets hosting stars C/O ratios cluster around the solar (0.55) value, and very few stars (if any) have C/O larger than 0.8. Interestingly, this is the value beyond which the carbon and oxygen chemistry will start changing dramatically, where (at high temperatures) CH$_4$ will replace CO as the most stable carbon bearing specie \citep{madhu2}. Since most stars seem to have a C/O ratio below this value, this phenomenon is not expected to take place in these systems. Therefore, we can test the effect of high stellar C/O ratio on the chemical compositions of Hot-Jupiters using our model for stars with C/O up to 0.8 without modifying the chemistry of the disk. \\
Results are shown in Fig. \ref{fig:jup2} for stellar C/O=0.8 and solar nebula refractory/volatile ratio. Since the chemistry of this system should be the same as in the solar C/O ratio case, the resulting planetary population distribution should be identical to that case, with the planets having higher C/O ratios scaled linearly by a factor of 1.45, as seen in Fig. \ref{fig:jup2}. The most interesting population in this case is the Hot-Jupiters forming in-situ (inside the water snowline), with significant ($\geq$ 60\%) core erosion. We can notice that these planets have a C/O ratio of higher than or equal to unity. This is the only population found in this work with C/O ratios higher than 0.9, and thus are expected to have a drastically different atmospheric chemistry, making them easily distinguishable with the next generation of telescopes.

\subsection{Effects of the envelope's opacity}
\begin{figure*}
	\includegraphics[scale=0.30]{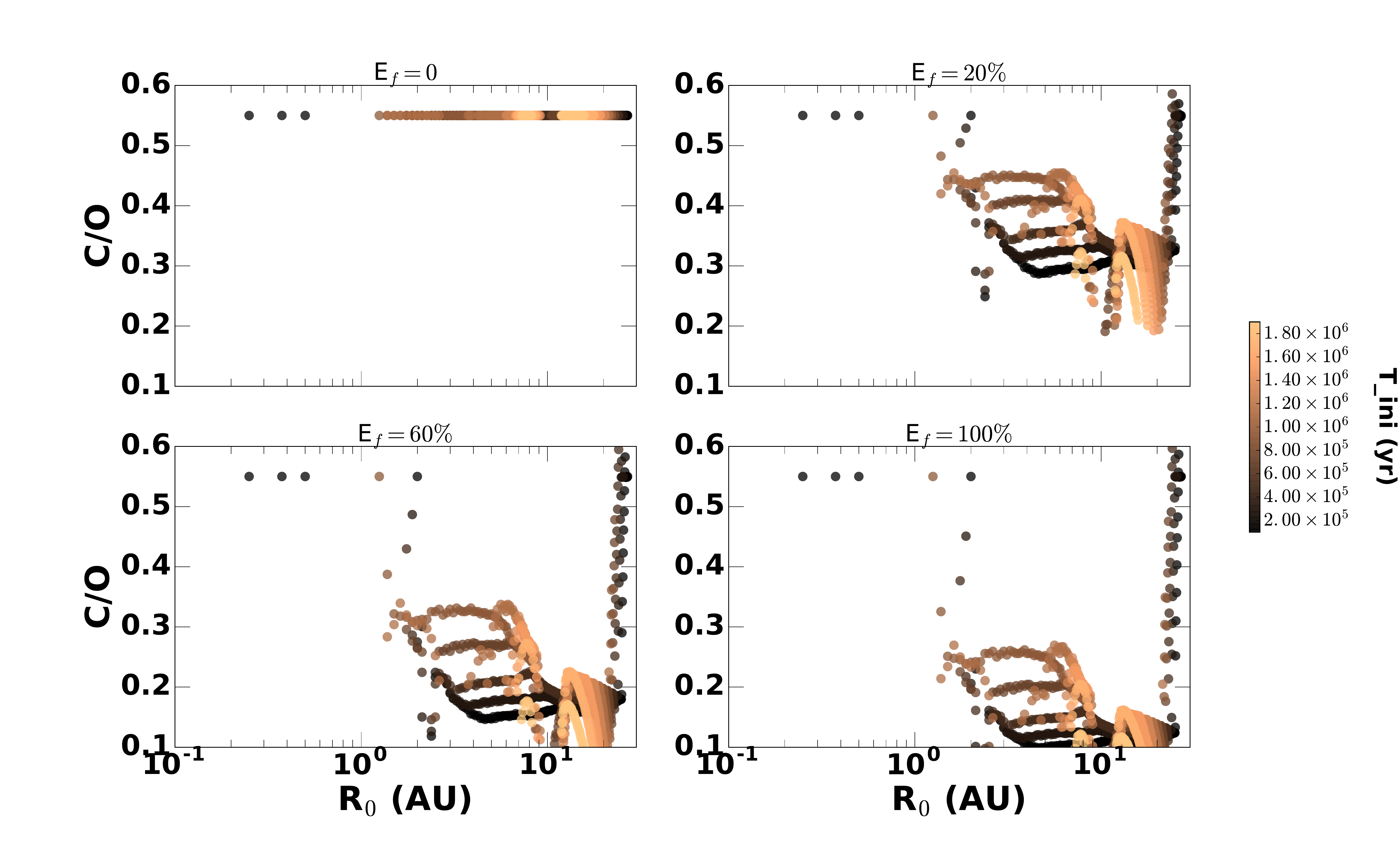}
   \caption{Same as Fig. \ref{fig:noref} but with all elements in purely volatile phases and $\kappa_{env}$ = 0.15. }
    \label{fig:noref2}
\end{figure*}

Now we study the effect of the envelope's opacity $\kappa_{env}$ on the C/O distribution of Hot Jupiters. The main effect of the $\kappa_{env}$ is to regulate the cooling ratio of the planet during the slow gas accretion phase, allowing it to contract more gas. Higher opacities will slow down the cooling rate and hence increase the time needed by the planet to reach the final fast hydrodynamical collapse phase where it transforms into a Jupiter mass planet. We hence run simulations with carbon and oxygen in purely volatile phases (with carbon in CO) but with $\kappa_{env}$ = 0.15 (a factor 3 higher than for our nominal cases). \\
The results are shown in Figure \ref{fig:noref2}. The difference between the cases with low and high envelope opacities are subtle. We notice that indeed higher $\kappa_{env}$ will slow down the giant planets formation as seen from the significantly lower number of Hot-Jupiters formed in this case (specially inside the water iceline where only a handful of Jupiters can now form in-situ). The general distribution trend however remains the same.

\subsection{Effects of Photoevaporation}
Another mechanism that can affect the disks and planets chemistry is disk photoevaporation \citep{guillot2006, monga2015}. Photoevaporation is in nature hydrodynamical, and hence not sensitive to the chemical composition of the gas. Therefore both carbon and oxygen will probably be affected due to PE at the same rate. The C/O ratio should therefore remain unaltered by the PE in a simple model where the gaseous volatiles are maintained in the disk through trapping in ices. Disk PE however can alter the carbon abundance - core mass correlation discussed above \citep{ali-dibc}.

\section{Caveats \& Future improvements}
\begin{itemize}
{\item The disk viscous evolution was not included in this work. It can affect the results indirectly via either its impact on the disk thermal structure (and thus the icelines locations) \citep{ali-dibch}, or through volatiles transport (by directly affecting the C/O ratio) \citep{ali-dibb}. Including these effects in future models is crucial for a more complete picture.
\item Core erosion is assumed to play a fundamental role in a planet's chemical composition. This however necessitate an efficient erosion mechanism and the planet's interior to be convective. Including more detailed models for these effects can give valuable insights. 
{\item  In the model we artificially reduced the pebbles mass flux and dominant size inside the snowline. Including these effects self consistently into the models is important to understand the evolution of planets in the inner most parts of the disk. }
}
\end{itemize}

\section{Summary \& Conclusions}
In this work we tried to link the chemical composition of Hot-Jupiters to their formation location using pebble accretion based population synthesis model with simplified chemistry. Our results can be summarized as following:
\begin{itemize}
\item Hot-Jupiters can form inside the water snowline via pebble accretion if their seed is injected early enough in the disk. This channel is significantly less efficient than forming them outside of the snowline. {This is due to the lower isolation mass in these regions leading to slow gas accretion, but also to the core formation via pebbles accretion taking more time due to the smaller pebbles size.}
\item Hot-Jupiters forming in-situ (inside the water snowline) will mostly have a higher C/O ratio than those forming outside of the snowline via disk migration. The exact difference between the two scenarios however significantly depend on the carbon and oxygen disk chemistry and amount of solid core erosion. The C/O ratios will vary more drastically in disks with significant amounts of C and O in volatile forms, in contrast with the solar system composition case where the variations are mild.
\item The only possible channel to form a Hot-Jupiter with a C/O higher than unity is if the stellar's C/O ratio is around 0.8. The highest possible C/O ratio in the nominal solar system case is 0.75, and 0.55 in the purely volatiles case. Planets forming between the 2 icelines in the purely volatile case can have C/O ratio as low as 0.2.
\item {Assuming the solar system chemistry case where most of the carbon is in refractories, the carbon abundance should correlate directly to the core mass in the Hot-Jupiters, and thus these two quantities can be used as a tracers for each other. Both Carbon and Oxygen abundances however are degenerated for the same core mass value due to the two possible formation scenarios leading in some cases to the same cores masses but with different chemical compositions.}  
\end{itemize}

\renewcommand\arraystretch{1.2}
\begin{table*}
\begin{center}
\caption{Initial conditions the disks and planets.}
\footnotesize
{\begin{tabular}{lcccc}
\hline
\noalign{\smallskip}

Parameter			& Range			& Step	\\
\hline
T$_{ini}$			& 10$^5$ - 2.0$\times$10$^6$ yr		& 10$^5$	 yr	 \\
R$_0$	& 0.125 - 37.5 AU		& 0.125 AU				 \\		
E$_f$				& 0 - 100\%  	& 20\%	\\		
\hline
M$_0$				& 1 $\times$ 10$^{-4}$ M$_\oplus$		& -		\\	
metal & 0.5 \% & - \\
Z$_0$ & 1 \% & -  \\
$f$  & 0.2 & - \\
$\kappa_{env}$ & 0.05 cm$^2$ g$^{-1}$ & - \\
$\rho_c$ & 5.5 g cm$^{-3}$& - \\
\hline	
\end{tabular}}\\

\label{t1}
\end{center}
\end{table*}

\renewcommand\arraystretch{1.2}
\begin{table*}
\begin{center}
\caption{Pebbles chemistry assumptions. }
\footnotesize

{\begin{tabular}{cccc}
\hline
Parameter & Inside H$_2$O IL (150 K) & Between H$_2$O and CO ILs & Outside CO IL (22 K)   \\
\hline
C (refractory) &  80\%  &  80\% &  80\%   \\
O (refractory) &  50\% &  50\% &  50\% \\
C (Ice)  & 0\% & 0\%  & 20\%   \\
O (Ice) & 0\% & 39\% & 50\% \\
C (Vapor)  & 20\%   & 20\%  & 0\% \\
O (Vapor) & 50\% & 11\%  & 0\% \\
\hline
Solar (Bulk disk) C/O & \multicolumn{3}{c}{0.55} \\
\hline
C/O refractories			& 0.88 & 0.88 & 0.88 	\\
C/O ices		& N/A & N/A & 0.22 	\\
C/O total solids			& 0.88 & 0.44 & 0.55	\\
C/O vapors & 0.22 & 1 & N/A 	\\

\hline

\end{tabular}}\\
{{Percentages are the fractions of Carbon and Oxygen in refractory, ice and gaseous component.} All percentages and ratios are of molar abundances. N/A means not defined. These are for the nominal (solar system chemistry) case only. }

\label{t2}
\end{center}
\end{table*}

\section*{Acknowledgements}

Data was analyzed using Python's Pandas package. We thank L. Kreidberg for reading and commenting on this manuscript. We thank B. Bitsch for providing us with the disk model fits, and answering our many questions. We thank the anonymous referee for useful comments that significantly improved the manuscript. Special thanks go to the Centre for Planetary Sciences group at the University of Toronto for useful discussions.








\appendix


\bsp	
\label{lastpage}
\end{document}